\documentstyle[11pt,Styles/newpasp,twoside,epsf]{article}
\def\edcomment#1{\iffalse\marginpar{\raggedright\sl#1\/}\else\relax\fi}
\reversemarginpar

\begin{document}
\title{Primordial Galaxy Formation and IGM Reionization}
 \author{Benedetta Ciardi}
\affil{MPA, Karl-Schwarzschild-Str. 1, 85741 Garching, Germany}        

\begin{abstract}
In this talk I will present a model for primordial galaxy formation. In
particular, I will review the feedback effects that regulate the process:
(i) radiative (i.e. ionizing and H$_2$--photodissociating photons) and (ii)
stellar (i.e. SN explosions) feedback produced by massive stars. I will
show the results of a model for galaxy formation and IGM reionization, which 
includes a self-consistent treatment of the above feedback effects. Finally,
I will describe a Monte Carlo method for the radiative transfer of ionizing 
photons through the IGM and discuss its application to the IGM reionization
problem.
\end{abstract}

\section{Introduction}

The application of the Gunn-Peterson test to QSOs absorption spectra suggests
that the Intergalactic Medium (IGM) is completely reionized by $z \sim 6$.
Several authors claim that the known population of quasars and galaxies
provides $\sim 10$ times fewer ionizing photons than are necessary to keep the
observed IGM ionization level; thus, additional sources of ionizing photons
are required at high redshift, the most promising being early galaxies
and quasars. Recent observational evidences suggest the
existence of an early population of pregalactic stellar objects which could
have contributed to the reionization and metal enrichment of the IGM.
In the following section, I will review the processes that regulate the formation
and the evolution of such galaxies.

\section{Primordial Galaxy Formation and Feedback Effects}

Once the gas has virialized into the potential well of preexisting dark
matter haloes, additional cooling is required in order for the gas to further
collapse and form stars. Objects with virial
temperatures $T_{vir}>10^4$~K rely on hydrogen Ly$\alpha$ line cooling for
their formation, while the main coolant for objects with $T_{vir} < 10^4$~K
({\it PopIII objects}) is molecular
hydrogen, a key species for galaxy formation in the early universe.
H$_2$ is a very fragile molecule, as it is
easily dissociated by photons with energy in the range 11.26-13.6~eV. Thus, the
existence of an UV flux below the Lyman limit, capable of dissociating
H$_2$ in collapsing structures can strongly influence the
formation of PopIII objects. Ciardi et al. (2000) have derived
the minimum mass that a forming galaxy requires to be able to shield
against an external incident dissociating flux: they found that the mass
is in the range $10^5 M_\odot<M<10^9 M_\odot$, depending on the intensity of
the incident flux and the formation redshift.

Once an object has collapsed, the first episode of star formation can strongly
influence the subsequent star formation process via the effects of mass
and energy deposition due to massive stars.
These processes may induce different phenomena, such as a heating and a loss
of the interstellar gas, or a dissociation of H$_2$ molecules in
star forming regions. As a consequence, the star formation rate decreases,
more or less dramatically depending on the mass of the object.

\section{IGM Reionization}

\begin{figure}
\plotfiddle{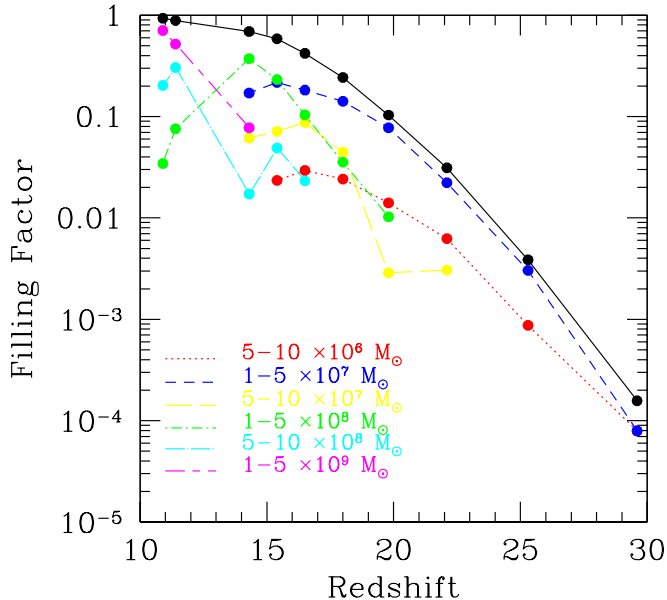}{2.80in}{0}{52}{52}{-160}{-110}
\vskip .5truecm
\caption{Evolution of HII filling factor as a function of redshift
(solid curve). The contribution from objects with masses in different
range is also shown: $5-10 \times 10^6 M_\odot$ (dotted line),
$1-5 \times 10^7 M_\odot$ (short dashed), $5-10 \times 10^7 M_\odot$
(long dashed), $1-5 \times 10^8 M_\odot$ (short dashed-dotted),
$5-10 \times 10^8 M_\odot$ (long dashed-dotted), $1-5 \times 10^9
M_\odot$ (short-long dashed).}
\end{figure}

Ciardi et al. (2000) have studied the interplay between the
primordial galaxy formation
process and the IGM reionization. In their analysis the authors have
included a self-consistent treatment of the feedback effects that
regulate the formation and the evolution of galaxies, as explained in
the previous section. One of the output of their model is the redshift
evolution of the HII filling factor, $f_{HII}$ (Fig.~1).
At $z>15$ the main contribution to $f_{HII}$ comes from small mass
objects ($M\sim  10^7 M_\odot$), while at lower redshift, when
the formation and evolution of such objects is suppressed by feedback
effects, the main contribution comes from galaxies of $M> 10^9
 M_\odot$. Thus, the contribution of small mass objects to the ionizing
photon budget is negligible. These galaxies
are nevertheless important as sinks of radiation on small scales.

To improve on the mentioned study, the assumption of constant IGM density
should be released and a realistic density distribution adopted.
In this case, the main complication would be to properly follow the
propagation of ionizing photons
into the IGM. To this aim, Ciardi et al. (2001) have developed the code {\tt CRASH},
a Monte Carlo approach to the radiative transfer of ionizing photons
produced by a source with known emission properties through a given
density field.

\begin{figure}
\plottwo{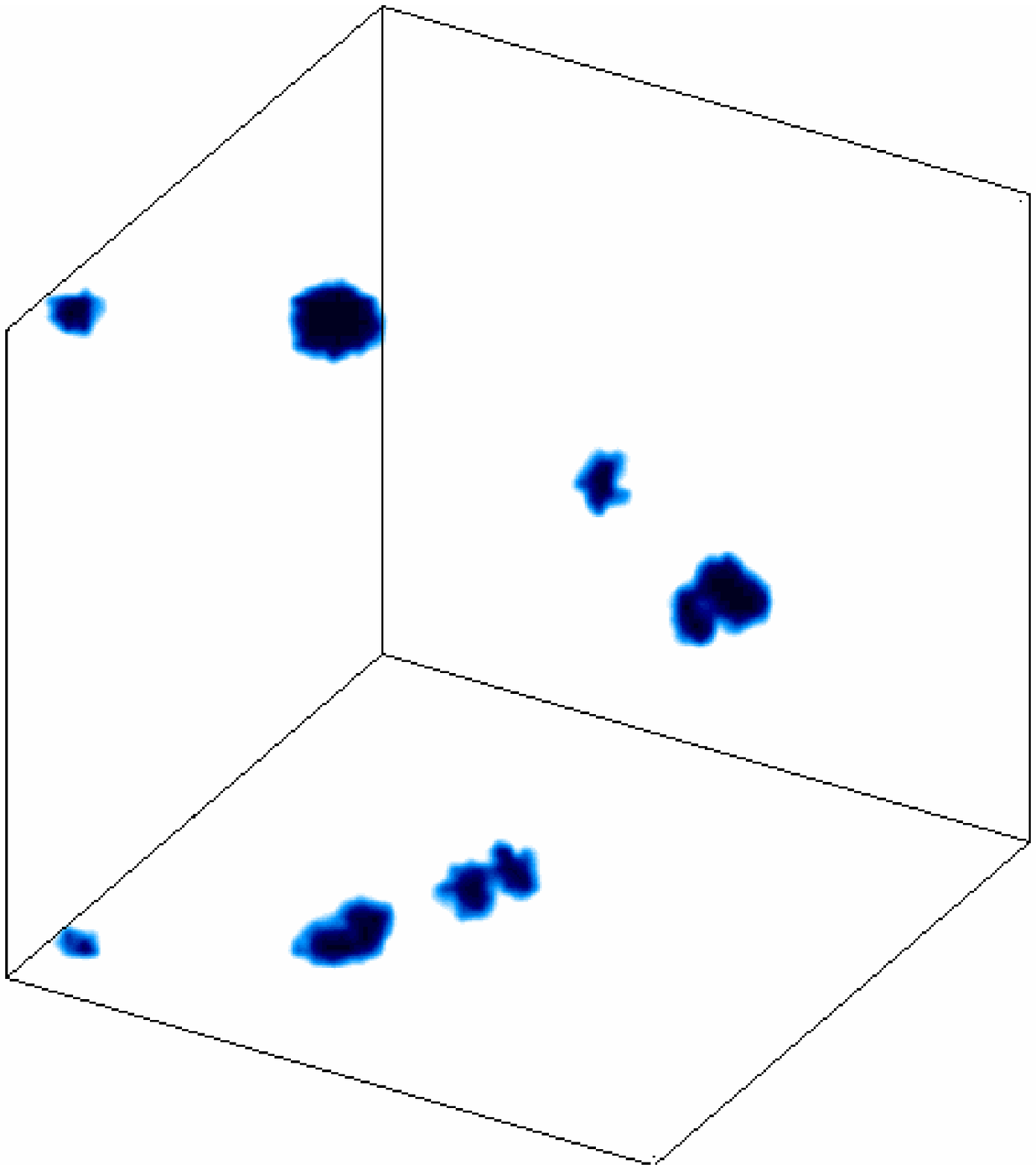}{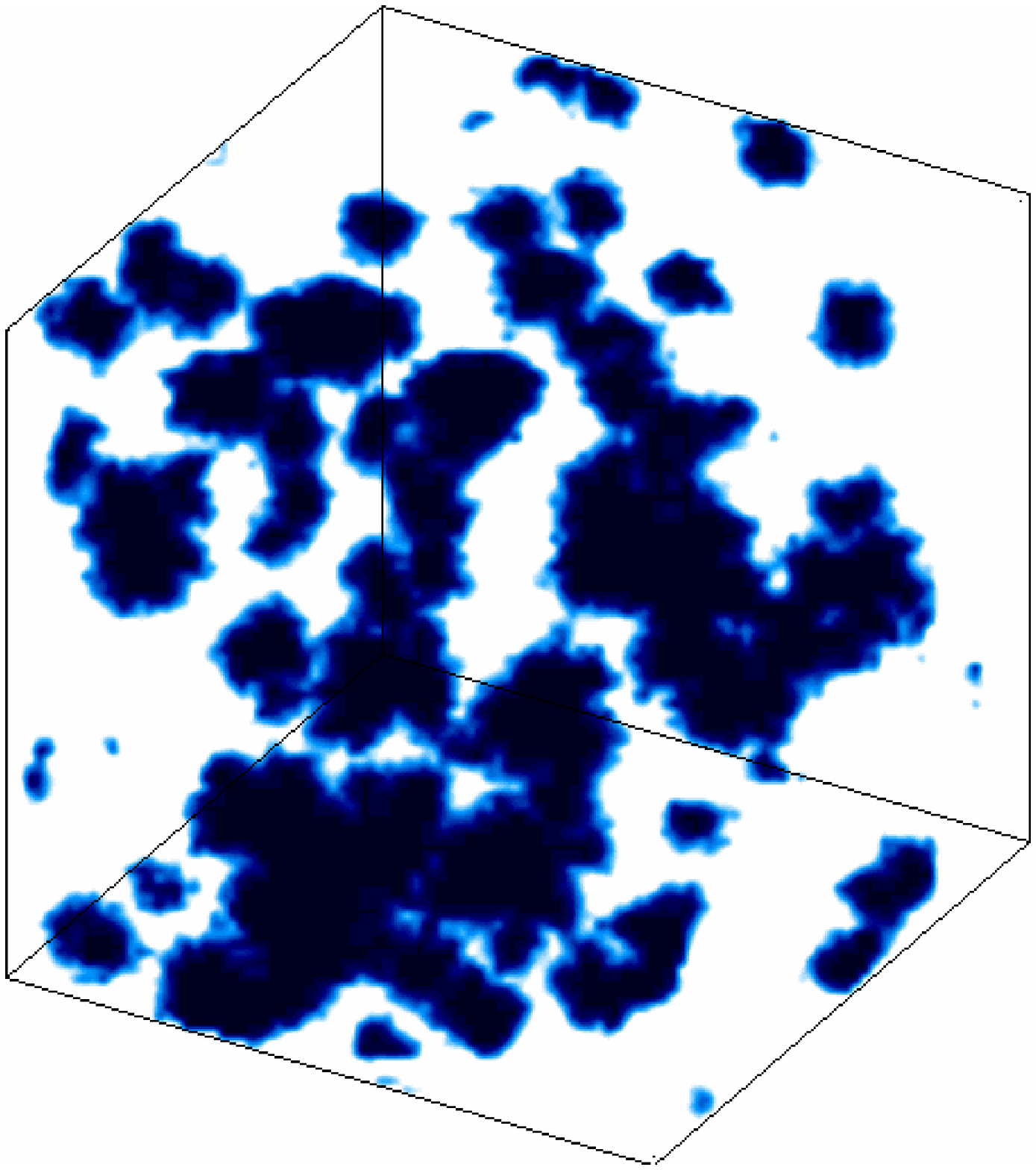}
\caption{Ionized regions in a 30 Mpc comoving simulation box at
$z\sim 13$ (left panel) and $z\sim 10$ (right panel).}
\end{figure}

Using {\tt CRASH},
we simulate the reionization process in a box of 30~Mpc comoving
(St\"{o}hr et al.; Ciardi, St\"{o}hr \& White in prep.). This choice of
the box dimension allows us to study a region large
enough to be representative of the universe
and at the same time to resolve objects with
masses $\sim 10^9 M_\odot$, the ones responsible for the
main ionizing photon production (see Fig.~1). The box is obtained
from a combination
of a high-resolution $\Lambda$CDM N-body simulation (Yoshida, Sheth
\& Diaferio 2001) and a semi-analitycal model of galaxy formation
(Springel et al. 2001).
In Fig.~2 the ionized regions produced by the simulation are shown,
at $z\sim 13$ (left panel) and $z\sim 10$ (right panel), when the
volume averaged ionization fraction is 0.1\% and 4\% respectively.

The reionization process critically depends on the choice of the
source emission properties. Here, we have adopted a Larson IMF and
a spectrum typical of PopIII stars; these assumptions give an upper
limit to the ionizing photon production. More critical is the
choice of the escape fraction of ionizing photons,
$f_{esc}$, for which we have adopted the indicative value of 20\%.
As $f_{esc}$ is a highly uncertain parameter,
Ciardi, Bianchi \& Ferrara (2001) have
tried to constrain it for a Milky Way type galaxy via 3D
numerical simulations, using the code {\tt CRASH} to follow the
photon propagation.
They have  compared results for a smooth Gaussian density
distribution and an inhomogeneous, fractal one, with realistic assumptions
for the ionizing stellar sources based on available data in the
solar neighborhood. In Fig.~3, the evolution of $f_{esc}$ as a function of
the total ionization rate, $\dot{\cal N}_\gamma$, is shown in the
case of a Gaussian (filled triangles) and a fractal (open circles)
density distribution. Values in the range $2\%<f_{esc}<50\%$ are found
depending on $\dot{\cal N}_\gamma$ and the density field, confirming
the high uncertainty of $f_{esc}$.

\begin{figure}
\plotfiddle{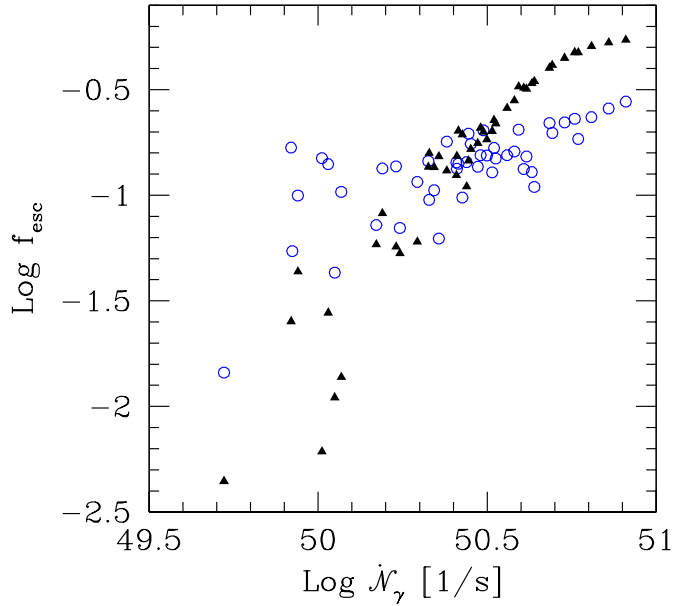}{2.80in}{0}{52}{52}{-160}{-110}
\vskip .5truecm
\caption{Evolution of the escape fraction as a function of the
total ionization rate for a Gaussian (filled triangles) and a fractal
(open circles) density distribution.}
\end{figure}

\section{Conclusions}

To properly describe the IGM reionization process, the following
ingredients are needed: i) a reliable model for structure formation, including
a self-consistent treatment of feedback effects and the best possible
balance between large volume and high resolution of the simulation;
ii) an accurate treatment of the radiative transfer of ionizing photons;
iii) a better constraint on the parameters involved in the calculation, such
as the source emission properties and the escape fraction.

\acknowledgments
I would like to thank all my collaborators in the projects.


\begin{references}
\reference Ciardi, B., Bianchi, S. \& Ferrara, A. 2001, MNRAS, in press
\reference Ciardi, B., Ferrara, A., Governato, F. \& Jenkins, A. 2000, MNRAS, 314, 611
\reference Ciardi, B., Ferrara, A., Marri, S. \& Raimondo, G. 2001, MNRAS, 324, 381
\reference Springel, V. et al. 2001, MNRAS, 328, 726
\reference Yoshida, N., Sheth, R. K. \& Diaferio, A. 2001, MNRAS, 328, 669
\end{references}
\end{document}